\documentclass[%
 reprint,
 amsmath,amssymb,
 aps,
]{revtex4-1}

\usepackage{graphicx}
\usepackage{dcolumn}
\usepackage{bm}

\usepackage{lipsum}
\usepackage{url}
\usepackage{hyperref}
\usepackage{natbib}
\usepackage{url}
\usepackage{multirow}
 \usepackage{mathtools}

\newcommand{\ff}{\textsc{f}}
\makeatletter
\newcommand*{\rom}[1]{\expandafter\@slowromancap\romannumeral #1@}
\makeatother

\begin{document}

\preprint{APS/123-QED}

\title{Noncommutative dispersion relation and mass-radius relation of white dwarfs}

\author{Arun Mathew}
\email{a.mathew@iitg.ac.in}
\author{Malay K. Nandy}%
 \email{mknandy@iitg.ac.in}
\affiliation{Department of Physics, Indian Institute of Technology Guwahati, 781039 India.
}%

\date{\today}

\begin{abstract}
The equation of state of the electron degenerate gas in a white dwarf is usually treated by employing the ideal dispersion relation. However, the effect of quantum gravity is expected to be inevitably present and when this effect is considered through a non-commutative formulation, the dispersion relation undergoes a substantial modification. In this paper, we take such a modified dispersion relation and find the corresponding equation of state for the degenerate electron gas in white dwarfs. Hence we solve the equation of hydrostatic equilibrium and find that this leads to the possibility of the existence of excessively high values of masses exceeding the Chandrasekhar limit although the quantum gravity effect is taken to be very small. It is only when we impose the additional effect of neutronization that we obtain white dwarfs with masses close to the Chandrasekhar limit with nonzero radii at the neutronization threshold. We demonstrate these results by giving the numerical estimates for the masses and radii of $\prescript{4}{2}{\mathbf{He}}$ and $\prescript{12}{6}{\mathbf{C}}$, and $\prescript{16}{8}{\mathbf{O}}$ white dwarfs.
\end{abstract}

\maketitle

\section{Introduction}\label{intro} 

The well-known Chandrasekhar limit for white dwarfs play a significant role in the present day understanding of astronomical observations. In particular, this mass limit has been a very important tool to characterize type Ia supernova that have served as stanadized candles in the measurement of their distances, and in particular in concluding about the accelerated expansion of the universe \cite{Riess1998, Perlmutter1999}. In determining the Chandrasekhar mass limit, classical gravity is employed and it becomes an important question whether this mass limit is modified because of the inevitable presence of the effects of the quantum gravity.

It is thus important to study the hydrostatic equilibrium of white dwarfs when the effect of quantum gravity is included in the description. In a simple formulation, it has been shown that quantum gravity leads to a generalized uncertainty relation $[\hat{x}_i,\hat{p}_j]=i\hbar \delta_{ij}(1+\beta \hat{\textbf{p}}^2)$\cite{Maggiore1993, Kempf1995, Pedram2012a} where $\hat{x}_i$ and $\hat{p}_j$ are the position and momentum operators and $\beta$ is a parameter due to the effect of quantum gravity. Since $\beta$ is a small parameter, this uncertainty relation will be effective in the high momentum region (such as the center of a massive white dwarf where the Fermi momentum is high).  As this uncertainty relation is different from the Heisenberg uncertainty principle $[\hat{x}_i,\hat{p}_j]=i\hbar \delta_{ij}$, the electron degenerate gas in a white dwarf is expected to be affected by it leading to a change in the equation of state. This problem was analyzed in detail \cite{Mathew2018} and it was found that white dwarfs with excessively high values of masses beyond the Chandrasekhar limit could be supported although the parameter $\beta$ is very small. However, when the condition of neutronization was imposed together with a feasible small value of $\beta$, a mass value close to the Chandrasekhar mass was obtained.

Since there are various alternative descriptions of quantum gravity, it becomes a natural question whether the above feature is preserved in the alternative formulations. It is thus important to analyze the problem in an alternative perspective of quantum gravity. The works of \cite{Amelinocamelia2000} and \cite{Amelinocamelia2002} in this direction suggested that the dispersion relation is substantially modified from the ideal case due to the effect of quantum gravity through noncommutativity where the space and time coordinates are treated as noncommuting quantities, such as $[\hat{x}_i,\hat{t}]=i\lambda \hat{x}_i$ and $[\hat{x}_i,\hat{x}_j]=0$, where $\lambda$ is a parameter due to the effect of quantum gravity. A special form of the modified dispersion relation, $E^2=p^2c^2(1+\lambda E)^2+m^2c^4$, was considered by \cite{Bertolami2010}, where $E$, $p$ and $m$ are the energy, momentum and mass of an electron. Since this dispersion relation is different from the ideal dispersion relation, $E^2=p^2c^2+m^2c^4$, we expect a modification in the equation of state of the electron degenerate gas in a white dwarf. This is expected to alter the stability of the star. Consequently in this paper, we take this modified dispersion relation to find the equation of state of the degenerate electron gas in a white dwarf and hence we solve the equation of hydrostatic equilibrium. We find that white dwarfs of excessively high values of masses beyond the Chandrasekhar mass are supported although the quantum gravity parameter $\lambda$ is taken to be very small. It is only when we impose neutronization into the problem, we get the mass values close to the Chandrasekhar limit. It is thus evident that whichever way we take the effect of quantum gravity, we reach the same conclusion: unbounded mass limits for white dwarfs although the effect of quantum gravity is taken to be very small. Since the effect of quantum gravity is inevitably present,  it is neutronization that is responsible for the limiting mass being nearly the Chandrasekhar mass.

We note that the effect of noncommutative dispersion relation was considered earlier for white dwarfs by \cite{Camacho2006} and \cite{Gregg2009}. The latter study reported slight increase or decrease in the limiting mass depending on the sign of the parameter $\lambda$, whereas the alternative approach via the generalized uncertainty relation formalism predicted an unbounded increase in mass and radius \cite{Rashidi2016}. On the other hand, in this work we analyze the effect of the modified dispersion relation in detail revealing features such as the possibility of excessively high values of masses of white dwarfs and the effect of neutronization that limits the mass of the white dwarf as indicated in the previous paragraph. 

In the present scenario, we further note that the behavior of the density $\rho$ with respect to the Fermi momentum $p_\ff$ remains the same as in the ideal case ($\rho\sim p_\ff^3$) and the noncommutative dispersion relation has no effect on it. This feature is quite unlike the behavior found with the generalized uncertainty relation where the density approaches a constant value as $p_\ff\rightarrow\infty$. Despite this disparity in the two approaches, we still find in the present case of noncommutative dispersion relation that white dwarfs can acquire arbitrarily high values of masses and radii similar to that in the case of generalized uncertainty relation. However, when we consider the role of neutronization together with a feasible value for the parameter $\lambda$, we find that the maximum possible masses for different white dwarfs (for example Helium, Carbon and Oxygen) are close to the Chandrasekhar limits.

The outline of the paper is as follows. In Section \ref{EOS}, we derive the equation of state for a degenerate electron gas where we also analyze the related asymptotic behaviors. In Section \ref{WD}, we consider the equations of hydrostatic equilibrium and discuss its asymptotic and exact solutions. In Section \ref{N}, we consider the limitation due to neutronization. Finally, we present a discussion and conclude the paper in Section \ref{conclusion}.

\section{Noncommutative Equation of State}\label{EOS} 
In this section, we obtain the number density $n$ and pressure $P$ of a degenerate electron gas employing the modified dispersion relation.  The asymptotic behavior of pressure $P$ in the limits of low and high Fermi momenta are also analyzed. 

\subsection{Modified Thermodynamic Behavior}\label{equation of state}
We employ the grand canonical ensemble (\cite{Landau1969}) for the electron gas, for which the grand potential can be expressed as $\Omega= -PV$ and 
\begin{equation}
\small
\Omega= -\frac{k_B TV}{\hbar^3\pi^2} \int_0^\infty  dp \ p^2 \ln \left[1 + \exp\left\{-\frac{(E_{\mathbf{p}}-\mu)}{k_B T}\right\}\right]. 
\end{equation}
The pressure $P$ can be immediately obtained from the above integral. In adiition, the identity $N = -\partial \Omega/\partial \mu$ gives the number density as 
\begin{equation}
\small
n = \frac{N}{V}=\frac{1}{\hbar^3\pi^2} \int_0^\infty  dp \ p^2 \left\{ \exp\left\{ \frac{E_{\mathbf{p}}-\mu}{k_B T}\right)+1\right \}^{-1}. 
\end{equation}

Since the electron gas in white dwarfs is completely degenerate to a very good approximation, we take the limit $T\rightarrow0$ in the above expressions and obtain
\begin{equation}\label{eq:Numberdensity}
n=\frac{1}{\hbar^3\pi^2}\int_0^{p_{\ff}} dp \ p^2,  \hspace{0.5cm}  P = \frac{1}{\hbar^3\pi^2} \int_0^{p_\ff} dp\  p^2 \left(E_\ff- E_{\mathbf{p}}\right),
\end{equation}
where $p_\ff$ is the Fermi momentum and $E_\ff$ is the Fermi energy.

Since the number density $n$ given by Eq.~(\ref{eq:Numberdensity}) remains unaffected by the modified dispersion relation, noncommutativity has no effect on it and we obtain the same expression as in the ideal case. We rewrite it in terms of the dimensionless variable $\xi=p_\ff/m_e c$ to obtain $n(\xi) = m_e^3 c^3\xi^3/(3\hbar^3\pi^2)$ and hence the mass density
\begin{equation}\label{eq:density}
\rho(\xi)=\mu_e m_u n(\xi) = K\left(\frac{\mu_em_u}{m_ec^2}\right)\frac{\xi^3}{3},                                            
\end{equation}
with $\mu_e=A/Z$  the number of nucleons per electron, $K=m_e^4c^5/\hbar^3\pi^2$, and $m_u=1.6605\times10^{-27}$ kg is the atomic mass unit.

It has been shown that the noncommutative formulation of quantum gravity leads to a dispersion relation more complicated than the ideal one \cite{Amelinocamelia2000, Amelinocamelia2002}. \cite{Bertolami2010} employed a simplified form of the dispersion relation 

\begin{equation}
E_{\bf p}^2=\mathbf{p}^2c^2(1+\lambda E_{\bf p })^2+m^2c^4
\end{equation}
where the parameter $\lambda$ quantifies the effect of quantum gravity.

Unlike the number density, the  pressure $P$ is modified due to the modification in the dispersion relation. This dispersion relation can be rearranged to obtain 

\begin{figure*}
\includegraphics[width=15.0cm]{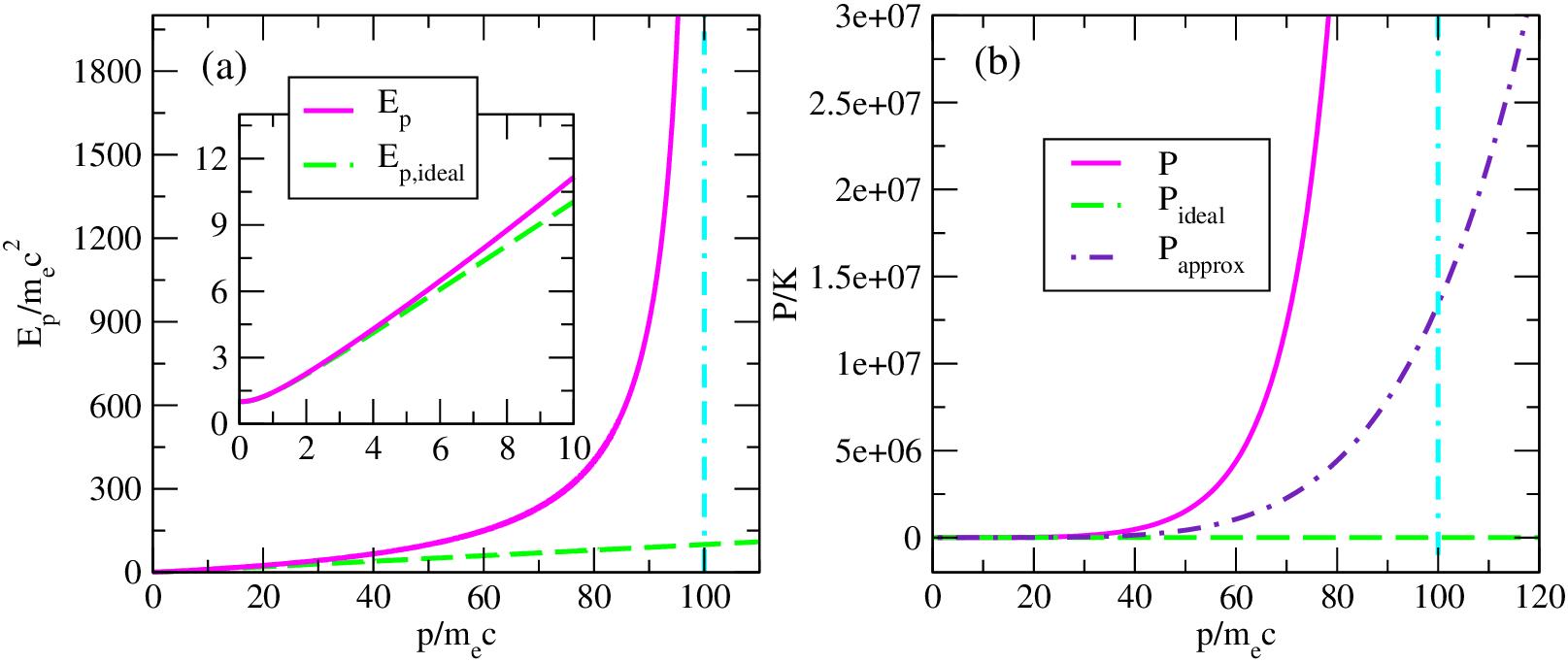}
\centering
\caption{(a) Noncommutative dispersion relation $E_{p}/m_e c^2$ (smooth curve) as a function of  $\tilde{p}=p/m_e c$ for the case $\alpha=0.01$ so that $\tilde{p}_{\rm max}=1/\lambda m_e c^2=1/\alpha=100$. Ideal dispersion $E_{p, \rm ideal}/m_e c^2$ (dashed curve) is also shown. The inset compares the two curves for low values of momentum. (b) Comparison of the noncommutative pressure $P$ (solid) with approximate $P_{\rm approx}$ (dot-dashed), and ideal $P_{\rm ideal}$ (dashed) expressions  given by Eqs.~(\ref{my_pressure}), (\ref{pressure_approx}) and (\ref{pressure_ideal}), respectively, for $p_{\ff \rm max}/m_e c=1/\alpha=100$.}
\label{Figure1}
\end{figure*}
\twocolumngrid
\begin{equation}\label{eq:dispersion}
E_{\bf p} = \frac{\lambda p^2c^2+\sqrt{p^2c^2+m^2c^4(1-\lambda^2 p^2 c^2)}} {1-\lambda^2 p^2 c^2}
\end{equation}
 This dispersion relation imposes a momentum cutoff at $p_{\rm max}=(\lambda c)^{-1}$ beyond which $E_{\bf p}$ becomes unphysical  (cf.~Figure~\ref{Figure1}a). We may rewrite it in terms of $\tilde{\textbf{p}}=\textbf{p}(m_e c)^{-1}$ as 
\begin{equation}\label{eq:f(y)}
f(\tilde{p}) = \frac{E_{\bf{p}}}{m_ec^2} = \frac{\alpha \tilde{p}^2+\sqrt{(1-\alpha^2)\tilde{p}^2+1}}{1-\alpha^2\tilde{p}^2}.
\end{equation}
where $\alpha=\lambda m_e c^2$. The behavior of  $f(\tilde{p})$ is shown in Figure~\ref{Figure1}a. For comparison, the ideal dispersion relation $E_{\bf{p,\rm ideal}}=\sqrt{p^2c^2+m_e^2 c^4}$ is also plotted in the same figure. We note that if we make the approximation $E_{\bf p} \approx \sqrt{p^2c^2+m^2c^4} + \lambda p^2c^2$ by neglecting the $O(\lambda^2)$ terms, the intrinsic momentum cut-off will be lost. We therefore avoid making this approximation to treat the high momentum region carefully. It may also be noted that this modified dispersion relation dictates the existence of a maximum density $\rho_{\rm max}=K \mu_em_u/(3m_ec^2\alpha^3)$ corresponding to the maximum cutoff in momentum $p_{\rm max}$. For example, for $\alpha=10^{-3}$, $\rho_{\rm max} = 1.9478\times10^{15}$ g cm$^{-3}$ and for $\alpha=10^{-4}$, $\rho_{\rm max} = 1.9478\times10^{18}$ g cm$^{-3}$.  

We obtain the pressure $P$ from Eq.~(\ref{eq:Numberdensity}) employing the complete noncommutative dispersion relation, given by Eq.~(\ref{eq:dispersion}), as 
\begin{widetext}
\small
\begin{equation}\label{my_pressure}
P(\xi)= K \left\{f(\xi) \int_0^\xi \tilde{p}^2 d\tilde{p} -\int_0^\xi f(\tilde{p})\tilde{p}^2 d\tilde{p} \right\} = K \left(f(\xi)\frac{\xi^3}{3}-g(\xi)\right) = K h(\xi)
\end{equation}
with
\begin{eqnarray}\label{eq:g}
 g(\xi)=  \frac{1}{\alpha^4}\left(2\tanh^{-1}{\alpha \xi}+\tanh^{-1}{\frac{\xi(1-\alpha^2)}{\alpha+\sqrt{1+(1-\alpha^2)\xi^2}}}-\frac{(2-\alpha^2)}{2\sqrt{1-\alpha^2}}\sinh^{-1}{\xi\sqrt{1-\alpha^2}}\right) \nonumber 
 - \frac{\xi}{3\alpha^3}\left(3+\alpha^2\xi^2+\frac{3\alpha}{2}\sqrt{1+(1-\alpha^2)\xi^2}\right).
\end{eqnarray}
\end{widetext}
The behavior of $P(\xi)$ is shown in Figure \ref{Figure1}b. We see that momenta higher than $\xi_{\rm max}$, determined by the cutoff $p_{\rm max}$ of the noncommutative dispersion relation (\ref{eq:dispersion}), are forbidden and the curve does not go beyond this limit.  In Figure 2, we compare the noncommutative equation of state given by Eqs.~(\ref{eq:density}) and (\ref{my_pressure}) with the ideal equation of state and the polytropic equations of state $P=K_n \rho^{1+1/n}$ with $n=3$ and $3/2$, where $K_{3}= \frac{1}{4}\left(\frac{3}{K}\right)^{1/3}\left(\frac{m_ec^2}{\mu_e m_u}\right)^{4/3}$ and $K_{3/2} =\frac{1}{5}\left(\frac{3}{K}\right)^{2/3}\left(\frac{m_ec^2}{\mu_e m_u}\right)^{5/3}$. The noncommutative equation of state clearly indicates that the density cannot exceed the maximum values $\rho_{\rm max}$ for different values of $\alpha$. This implies that the effect of quantum gravity forbids the star to have an infinite density (at the center).

 \begin{figure}
\centering
\includegraphics[width=8.5cm]{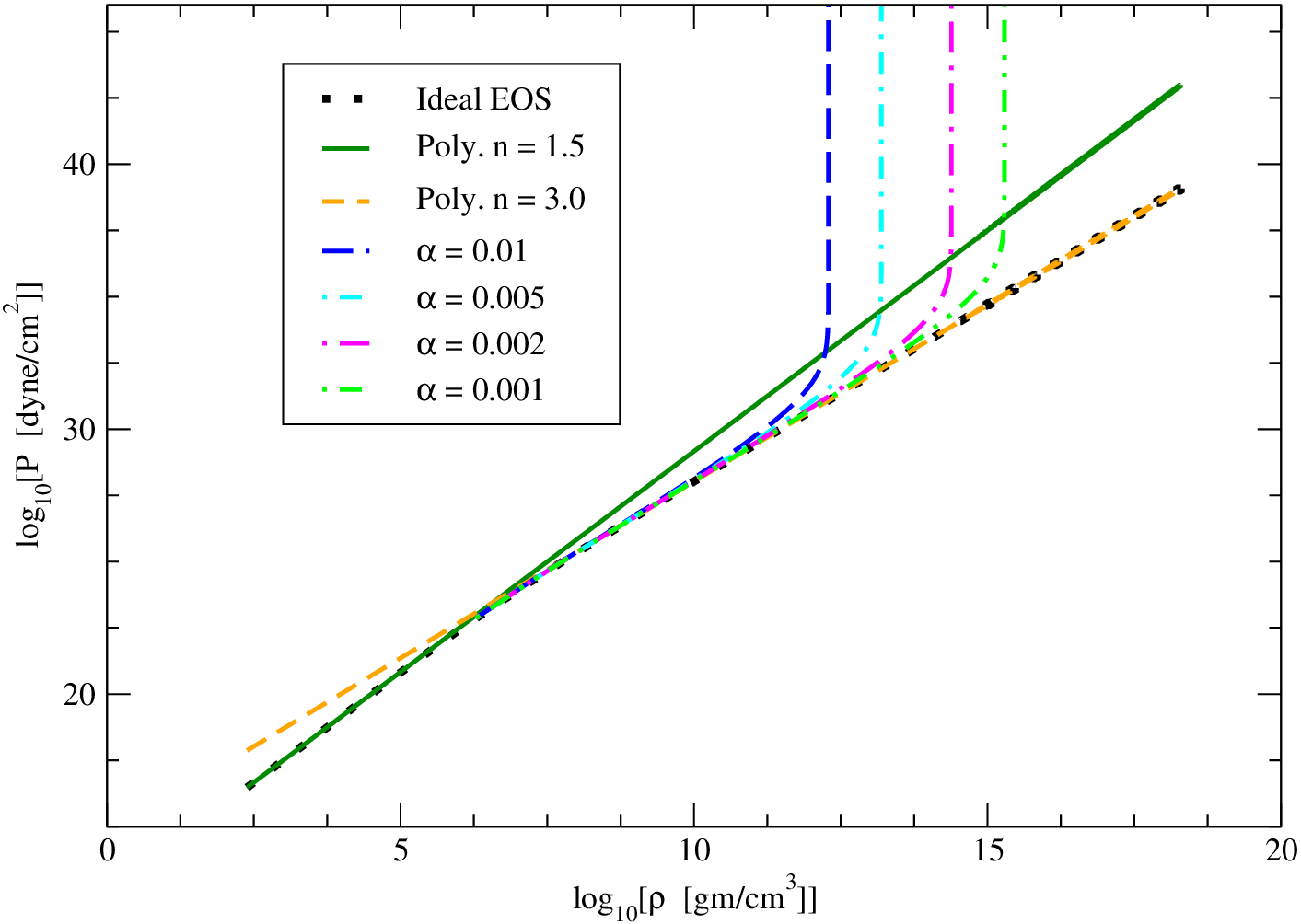}
\caption{Comparison of the noncommutative equation of state (for different values of $\alpha$) given by Eqs.~(\ref{eq:density}) and (\ref{my_pressure}) with the ideal equation of state given by Eqs.~(\ref{eq:density}) and (\ref{pressure_ideal}). Also shown are the polytropic equations of state $P=K_n \rho^{1+1/n}$ with $n=3/2$ and $3$.}
\label{Figure_After_1}
\end{figure}

Consistency of the equation of state connected by the above noncommutative expressions for $P(\xi)$ and $n(\xi)$ follows immediately as they satisfy the well-known thermodynamic relation $dP/d\mu=n$, where $\mu$ is the chemical potential. The left-hand side of this relation, in terms of the dimensionless parameter $\xi$, becomes 
\begin{equation}\label{eq:thermo}
\frac{dP}{d\mu}= \frac{K}{m_e c^2}   \frac{dh}{d\xi} \frac{1}{df/d\xi },
\end{equation}
where we have used $\mu=m_e c^2 f(\xi)$, which is the modified expression for the Fermi energy. The differentials in the above equation can be obtained from Eqs.~(\ref{eq:f(y)}), (\ref{eq:g})  and (\ref{my_pressure}) as 
\begin{equation}\label{eq:f'}
\small
\frac{df}{d\xi}= \xi \left\{\frac{(1+\alpha^2 \xi^2)+(1-\alpha^2\xi^2)\alpha^2+2\alpha\sqrt{1+(1-\alpha^2)\xi^2}}{(1-\alpha^2\xi^2)^2 \sqrt{1+(1-\alpha^2)\xi^2}}\right\}
\end{equation}
and   
\begin{equation}\label{eq:identity}
\frac{dh}{d\xi}=\frac{\xi^3}{3}\frac{df}{d\xi}.
\end{equation}
Using Eq.~(\ref{eq:identity}) in (\ref{eq:thermo}), it immediately follows that  $dP/d\mu= (K/3m_e c^2)\xi^3 = n$, ensuring consistency with the thermodynamic relation.

We thus see that, in noncommutative geometry, the equation of state undergoes a drastic modification due an intrinsic momentum cutoff inherent in the modified dispersion relation. This situation is quite unlike the scenario following from the generalized uncertainty principle where the equation of state undergoes a drastic modification due to a change in the measure of the phase space despite the dispersion relation remains ideal. A detailed analysis of this latter scenario is given in Ref. \cite{Mathew2018}.

\subsection{Ideal and Asymptotic Behaviors}\label{EOS_Asym} 

 It is easy to show that, in the limit $\alpha\rightarrow0$,  the parametric forms of the pressure corresponding to the ideal degenerate case can be recovered. The leading order terms in the expansion of Eq.~(\ref{my_pressure}) are obtained as  
\begin{equation}\label{pressure_ideal}
P_{\rm ideal}(\xi)=\frac{K}{24}\left\{\sqrt{1+\xi^2}(2\xi^3-3\xi)+3\sinh^{-1}{\xi}\right\}. 
\end{equation}

Figure~\ref{Figure1}b compares the noncommutative pressure $P(\xi)$ given by Eq.~(\ref{my_pressure}) with the ideal pressure $P_{\rm ideal}(\xi)$ given by Eq.~(\ref{pressure_ideal}). It may be noted that there is a large deviation between the two expressions for higher values of Fermi momentum. In the noncommutative case, the pressure increases faster and approaches infinity as the Fermi momentum approaches $p_{\rm max}$. This behavior is quite different from the ideal case where the energy density approaches infinity at a slower rate beyond $p_{\rm max}$. 

A first correction to the ideal  case can be obtained by a Taylor expansion about $\lambda=0$ and by retaining the $O(\lambda)$ term. In this approximation,  $E_{\mathbf p, \rm approx} = \lambda p^2c^2+\sqrt{p^2c^2+m^2c^4}$. The corresponding pressure turns out to be
\begin{widetext}
\begin{equation}\label{pressure_approx}
P_{\rm approx}(\xi)=K\left\{\frac{1}{24} \sqrt{1+\xi^2}(2\xi^3-3\xi)+\frac{1}{8}\sinh^{-1}\xi+2\alpha\frac{\xi^5}{15}\right\} = P_{\rm ideal} +  \frac{2}{15}K \alpha \xi^5,
\end{equation}
\end{widetext}
This approximate expression is also compared with the other cases in Figures \ref{Figure1}b. We see that the noncommutative momentum cutoff $p_{\rm max}$ (or $\xi_{\rm max} =\alpha^{-1}$) of the complete dispersion relation is not respected by the approximate expression $P_{\rm approx}(\xi)$ and it deviates strongly from the noncommutative expression $P(\xi)$. This indicates that the approximate form $P_{\rm approx}(\xi)$ is a not good approximation for high values of Fermi momentum near $\xi_{\rm max}$. This is due to the fact that the approximate dispersion relation given by $E_{\mathbf p, \rm approx}$ does not impose any restriction on the momentum values. On the other hand, the complete noncommutative dispersion relation $E_{\bf p}$ constrains momentum values by imposing a momentum cutoff $p_{\rm max}$. The importance of our present analysis lies in the fact that we use the complete noncommutative dispersion relation without making any approximations so that its basic feature of a maximum attainable momentum $p_{\rm max}$ is retained. 

We next analyze the asymptotic behavior of the noncommutative pressure $P(\xi)$ given by Eq.~(\ref{my_pressure}) in the low and high momentum limits, $\xi\rightarrow0$ and $\xi\rightarrow\ \xi_{\rm max}$. For low values of $\xi$, it is obtained as 
\begin{equation}\label{pressure_low}
P_{\rm low}(\xi) = K (1+2\alpha) \frac{\xi^5}{15}.
\end{equation}
It is important to note that, even in this limit, the effect of noncommutativity persists due to the presence of the term proportional to $\alpha$ at the order $\xi^5$. We shall see this feature to be present when we analyze the mass-radius relation for low values of central Fermi momentum $\xi_c$. Moreover, Eqs.~(\ref{eq:density}) and (\ref{pressure_low}) imply $P_{\rm low}\sim\rho^{5/3}$  which can be seen in Figure \ref{Figure_After_1} where the noncommutative and the ideal equations of state coincide in the low momentum region.

In the high momentum region $\xi\sim \xi_{\rm max}$, we expand the noncommutative expression for pressure $P$ assuming the momentum to be close to $p_{\rm max}$ (or $\xi_{\rm max}$), to obtain 
\begin{equation}\label{pressure_high}
P_{\rm high}(\xi)= \frac{K}{\alpha^4}  \left\{\frac{1}{1-\alpha \xi} - \ln\left(\frac{2\alpha^2}{1-\alpha \xi}\right) - C (\alpha)\right\}
\end{equation}
where 
\begin{equation}
C (\alpha) =  \tanh^{-1}\left(\frac{1-\alpha^2}{1+\alpha^2}\right) - \sinh^{-1}\frac{1}{\alpha}-\frac{11}{6} 
\end{equation}

Thus, when the central Fermi momentum $\xi_c$ is close to $\xi_{\rm max}=1/\alpha$, the central pressure approaches infinity. This boundless increase in the pressure for very high values of $\xi$ should be able to counteract gravitational pull in very massive white dwarfs. This feature will show up more explicitly later when we analyze the mass-radius relation for high values of the central Fermi momentum. Moreover, this feature can be seen in Figure \ref{Figure_After_1} where the pressure approaches infinity and the density approaches constant values $\rho_{\rm max}=K \mu_em_u/(3m_ec^2\alpha^3)$ as implied by  Eq.~(\ref{eq:density}). Unlike the ideal case, where $P_{\rm high}\sim \rho^{4/3}$, the behaviour is remarkably different in the high momentum region of the noncommutative equation of state.

\section{Noncommutative white dwarfs}\label{WD}

In this section, we obtain the mass-radius relation of Helium white dwarfs with the equation of state obtained in Section \ref{equation of state} from the noncommutative dispersion relation. In the framework of Newtonian gravity, the condition of hydrostatic equilibrium for a spherical distribution of matter is given by 
\begin{equation}\label{eq:HS1}
\frac{dP}{dr}=-\frac{G m(r)\rho(r)}{r^2}
\end{equation}
with
\begin{equation}\label{eq:HS2}
\frac{dm}{dr}=4\pi\rho(r)r^2.
\end{equation}

Combining Eqs.~(\ref{eq:HS1}) and (\ref{eq:HS2}), we get
\begin{equation}\label{eq:FHS}
\frac{1}{r^2}\frac{d}{dr}\left(\frac{r^2}{\rho}\frac{dP}{dr}\right)+4\pi G \rho(r)=0
\end{equation}

Substituting Eqs.~(\ref{eq:density}) and (\ref{my_pressure}) and using the dimensionless variable $x=r/R_0$ in Eq.~(\ref{eq:FHS}) yields

\begin{equation}\label{eq:mylaneemdeneq}
\frac{1}{x^2}\frac{d}{dx}\left(x^2 f'(\xi)\frac{d\xi}{dx}\right)+\frac{\xi^3}{3}=0
\end{equation}
where $R_0=(4\pi G K)^{-1/2}\left(m_e c^2/\mu_e m_u\right)=2242.77$ km.

\subsection{Asymptotic Solutions}\label{WD_Asym}

In the limit $\xi\rightarrow 0$, that is, for low values of $\xi$, it can be shown that $f'(\xi)=(1+2\alpha)\xi$. Thus Eq.~(\ref{eq:mylaneemdeneq}) can be rewritten as 
\begin{equation}\label{eq:}
\frac{(1+2\alpha)}{2}\frac{1}{x^2}\frac{d}{dx}\left(x^2 \frac{d\xi^2}{dx}\right)+\frac{\xi^3}{3}=0
\end{equation}

Now, taking $\xi^2(x)/\xi^2_c$ as $\theta(x)$, with $\xi_c$ the central dimensionless Fermi momentum, and defining a new dimensionless coordinate $\eta=\sqrt{2/3}\sqrt{\xi_c/(1+2\alpha)}\ x$, we reduce the above equation to
\begin{equation}
\frac{1}{\eta^2} \frac{d}{d\eta}\left(\eta^2\frac{d\theta}{d\eta} \right) +\theta^{3/2}=0
\end{equation}
which is the Lane-Emden equation of index $3/2$. The numerical solution for this differential equation is given in Weinberg \cite{Weinberg1972}.  For the boundary conditions $\theta(0)=1$ and $\theta'(0)=0$, one can immediately obtain the radius of the white dwarf as
\begin{equation}\label{eq:RLxi}
R = \sqrt{\frac{3}{2\xi_c}}\ (1+2\alpha)^{1/2}R_0\eta_R
\end{equation}
where $\eta_R=3.65375$ is the first zero of the Lane-Emden function $\theta(\eta)$ of index $3/2$. 

Similarly the asymptotic behavior of the mass of the white dwarf can be obtained from the integral expression of Eq.~(\ref{eq:HS2}), namely,
\begin{equation}\label{mass_equation}
M=4\pi\int_0^R \rho(r) r^2 dr = 4\pi K\left(\frac{\mu_em_u}{m_ec^2}\right) \int_0^R \frac{\xi^3}{3}  r^2 dr.
\end{equation}

Using $\tilde{M}=M/M_0$ and $\tilde{R}=R/R_0$ with $M_0=(4\pi K)^{-1/2}G^{-3/2} \left(m_e c^2/\mu_e m_u\right)^2=0.41659$ M$_\odot$ in the above equation, we get $\tilde{M}=\frac{1}{3}\int_0^{\tilde{R}}\xi^3 x^2 dx$. We rewrite this equation in the new dimensionless variable $\eta$, yielding 
\begin{equation}
\tilde{M}= \sqrt{\frac{3\xi_c^3}{8}} \ (1+2\alpha)^{3/2} \int_0^{\eta_R}\theta^{3/2} \eta^2 d\eta,
\end{equation}
thus obtaining the mass of the white dwarf as 
\begin{equation}\label{eq:MLxi}
M = -\sqrt{\frac{3\xi_c^3}{8}}  (1+2\alpha)^{3/2} M_0 \eta^2_R \left(\frac{d\theta}{d\eta}\right)_{\eta=\eta_R}
\end{equation}

The value of $\left(-\eta^2d\theta/d\eta\right)_{\eta=\eta_R}$ is $2.71406$ \cite{Weinberg1972}. Thus the above asymptotic analysis predicts that $R\sim(1+2\alpha)\xi_c^{-1/2}$ and $M\sim(1+3\alpha)\xi_c^{3/2}$ indicating the persistence of the effect of noncommutativity through the presence of the parameter $\alpha$ even for very low values of the central Fermi momentum. The presence of $\alpha$ (or $\lambda m_e c^2$) in these expressions suggests increase on the order of $\alpha$ in mass and radius of white dwarfs. We also note that for $\alpha=0$, the above mass-radius relation approaches the Chandrasekhar relation for low values of central Fermi momentum. 

On the other hand, in the limit $\xi\rightarrow\xi_{\rm max}=\alpha^{-1}$, $f'(\xi)=(1/\alpha^2)\left(\xi-1/\alpha\right)^{-2}$, so that Eq.~(\ref{eq:mylaneemdeneq}) reduces to 
\begin{equation}\label{eq:}
\frac{1}{x^2}\frac{d}{dx}\left(\frac{x^2}{(\xi-\frac{1}{\alpha})^2}\frac{d\xi}{dx}\right)+\frac{1}{3\alpha}=0
\end{equation}

Letting $\phi=\alpha/(1-\alpha\xi)$ yields 
\begin{equation}\label{eq:}
\frac{1}{x^2}\frac{d}{dx}\left(x^2\frac{d\phi}{dx}\right)+\frac{1}{3\alpha}=0
\end{equation}

Defining $(\phi(x)-\alpha)/(\phi_c-\alpha)=\theta(x)$, where $\phi_c=\alpha/(1-\alpha\xi_c)$ and redefining the dimensionless radius as $x=\sqrt{3\alpha (\phi_c-\alpha)}\zeta$, the above equation takes the form 
\begin{equation}\label{eq:}
\frac{1}{\zeta^2}\frac{d}{d\zeta}\left(\zeta^2\frac{d\theta}{d\zeta}\right)+1=0,
\end{equation}
which is the Lane-Emden equation of index zero whose numerical solution is already known. Thus the radius of the white dwarf is given by 
\begin{equation}\label{eq:RHxi}
R = \alpha \sqrt{\frac{3\alpha\xi_c}{1-\alpha\xi_c}} R_0\zeta_R
\end{equation}
where $\zeta_R=\sqrt{6}$ is the zero for the Lane-Emden function $\theta(\zeta)$ of index zero. 

The mass of the white dwarf can be obtained from Eq.~(\ref{mass_equation}) by taking the appropriate limit $\xi\rightarrow\xi_{\rm max}$ and using the above dimensionless coordinate $\zeta$, we get
\begin{equation}
\tilde{M}= \frac{1}{9} \left(\frac{3\alpha\xi_c}{1-\alpha\xi_c}\right)^{3/2}\zeta_R^3.
\end{equation}
Consequently the mass of the white dwarf is obtained as 
\begin{equation}\label{eq:MHxi}
M= \frac{M_0}{9} \left(\frac{3\alpha\xi_c}{1-\alpha\xi_c}\right)^{3/2}\zeta_R^3.
\end{equation}

Since Eqs.~(\ref{eq:RHxi}) and (\ref{eq:MHxi}) were obtained with the assumption of the central Fermi momentum $\xi_c$ approaching the maximum value  $\alpha^{-1}$, they  are valid near $\xi_{\rm max}$ (= $\alpha^{-1}$). In this limit the quantity $(1-\alpha \xi_c)$ approaches zero so that the mass and radius  diverge as $M\rightarrow(1-\alpha \xi_c)^{-3/2}$ and $R\sim(1-\alpha \xi_c)^{-1/2}$ as $\xi_c$ tends to $\xi_{\rm max}$. In fact, the largeness of the mass and radius will depend on how close is $\xi_c$ with respect to $\alpha^{-1}$. Thus, both mass and radius increase unboundedly as the central Fermi momentum $\xi_c$ approaches the maximum cutoff value $\xi_{\rm max}=\alpha^{-1}$. 

From Eqs.~(\ref{eq:RHxi}) and (\ref{eq:MHxi}) we obtain $MR^{-3}=$ Const. Since those expressions are valid for excessively high values of the Fermi momentum, this implies $M\sim R^3$. Since a solid sphere of uniform density has its mass proportional to its volume ($\sim R^3$), this suggests an approximately constant density in most part of the star. In an alternative description (\cite{Mathew2018}) based on the generalized uncertainty principle of quantum gravity, the same features were observed although with a completely different equation of state. 

\subsection{Exact Solutions}\label{WD_exact}

We employ the noncommutative equation of state obtained in Section \ref{equation of state}. Substituting Eq.~(\ref{eq:identity}), and using the definitions $m=M_0 u$ and $r=R_0 x$ in Eqs.~(\ref{eq:HS1}) and (\ref{eq:HS2}), we obtain
\begin{equation}\label{eq:FHS1}
\frac{d\xi}{dx}=-\frac{1}{f'(\xi)}\frac{u(x)}{x^2}
\end{equation}
and
\begin{equation}\label{eq:FHS2}
\frac{du}{dx}=\frac{1}{3}\xi^3x^2.
\end{equation}

\begin{figure*}[]
  \centering
  \includegraphics[width=15.0cm]{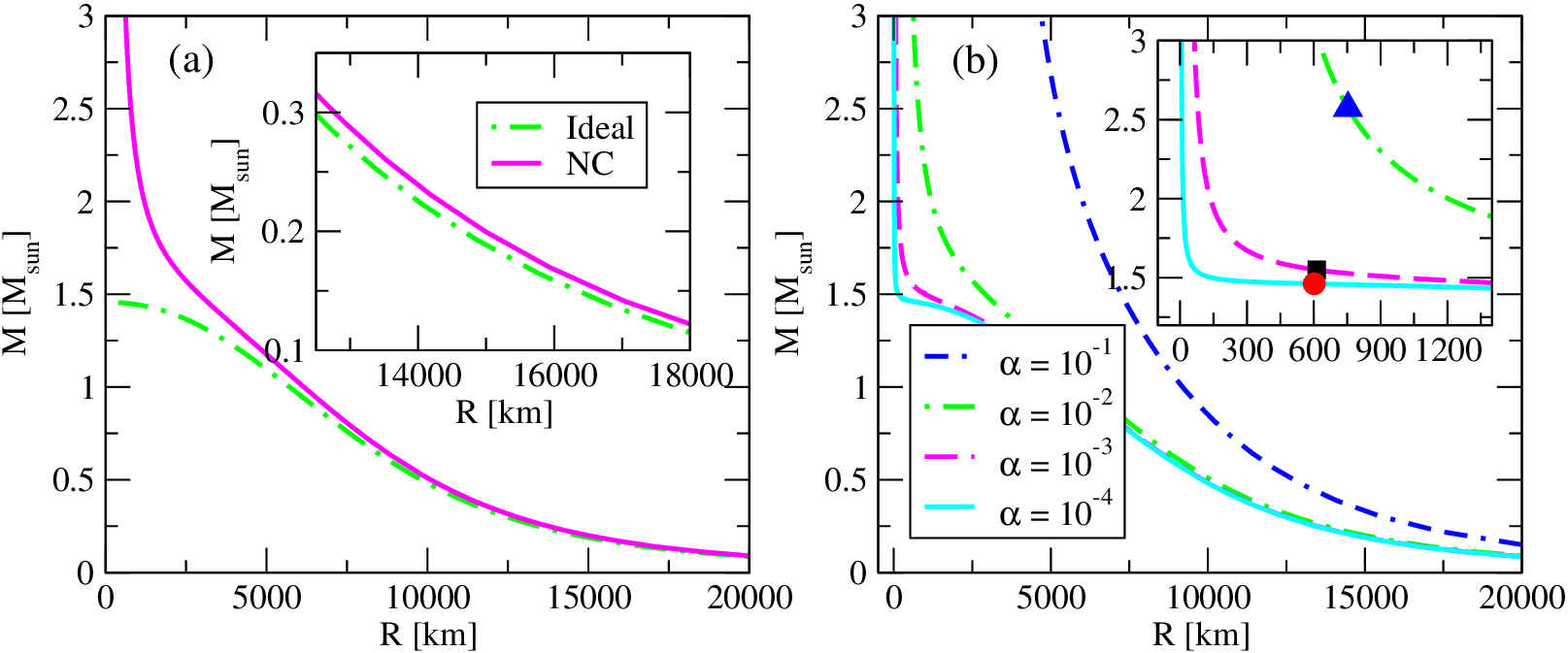}
  \caption{Mass-radius relations for Helium white dwarfs. (a) Solid curve (noncommutative equation of state with $\alpha=10^{-2}$) and dot-dashed curve (ideal equation of state). The inset shows slight departure of the noncommutative curve from the ideal curve for low $\xi_c$. (b) Plots with noncommutaive equation of state for different values of $\alpha$. The inset shows the behavior near the ``turning points" where the symbols represent the neutronization threshold points: $M=2.5734$ M$_\odot$, $R=753.24$ km (triangle) for $\alpha=10^{-2}$, $M=1.5495$ M$_\odot$, $R=613.28$ km (square) for $\alpha=10^{-3}$, and $M=1.4614$ M$_\odot$, $R=602.35$ km (circle) for $\alpha=10^{-4}$. In the inset of (b), the $x$-axis denotes the radius $R$ (in km) and the $y$-axis denotes the mass $M$ (in M$_\odot$).}
  \label{Figure2}
  \end{figure*}

The above two first-order differential equations are integrated simultaneously employing the fourth order Runge-Kutta method with the boundary conditions $\xi(0)=\xi_c$ and $u(0)=0$ until the surface defined by $\xi(\tilde R)=0$ is reached. The results of numerical integration are shown in Figures \ref{Figure2}a and \ref{Figure2}b. 
 
In Figure \ref{Figure2}a, we notice that for large central Fermi momenta $\xi_c$, the mass-radius relation of the noncommutative case departs considerably from the ideal (commutative) case. On the other hand, the two mass-radius curves come very close to each other (without coinciding) for lower values of the central Fermi momentum, as shown in the inset of Figure  \ref{Figure2}a. Our numerical data indicate an increase of about $3.2\%$ in the mass for a white dwarf of $0.17$ M$_\odot$ for $\alpha=10^{-2}$, whereas this increase is about $0.03\%$ for $\alpha=10^{-4}$.

It is important to note that, for small values of the central Fermi momentum, the mass-radius relation due to the noncommutative dispersion relation does not coincide with the ideal degenerate case, which is shown in the inset of Figure \ref{Figure2}a. This behavior can be seen from the asymptotically obtained mass and radius expressions given by Eqs.~(\ref{eq:MLxi}) and (\ref{eq:RLxi}). The persistence of the deformation parameter $\alpha$ even in the low momentum regime exhibits this disparity on the right-hand part of the mass-radius curve. This result leads to the implication that the value of the deformation parameter $\lambda$ due to the effect of quantum space-time fluctuations may possibly be observed via high precision measurements on naturally existing white dwarfs.

\begin{figure}
\centering
\includegraphics[width=8.5cm]{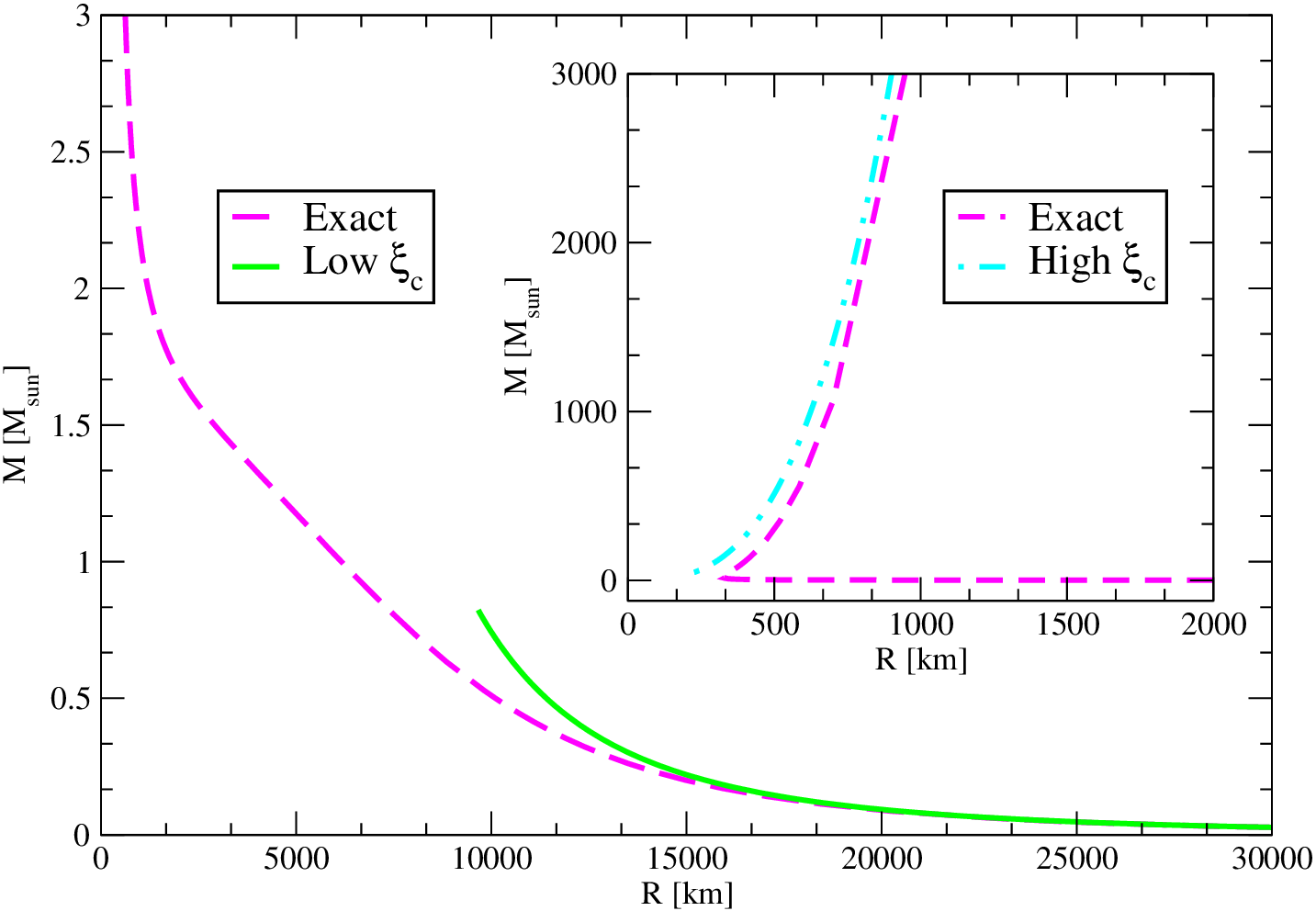}
\caption{Mass-radius curves for Helium white dwarfs with the noncommutative equation of state for $\alpha=0.01$. The dashed curve represents the exact solution. The solid curve represents the approximate solutions (\ref{eq:MLxi}) and (\ref{eq:RLxi}). The inset shows slight departure of the asymptotic solutions (\ref{eq:MHxi}) and (\ref{eq:RHxi}) from the exact solution.}
\label{Figure3}
\end{figure}

In Figure \ref{Figure2}b, we display the mass-radius relations with the noncommutative equation of state given by Eqs.~(\ref{eq:density}) and (\ref{my_pressure}) for different values of $\alpha$, namely, $\alpha=0.0001$, $0.001$, $0.01$, and $0.1$. We see that, for large values of $\alpha$,  the mass-radius relation departs from those of smaller values. This is expected since the effect of quantum fluctuations of space-time is expected to be stronger for large values of $\alpha$. Besides this, we also note that the Chandrasekhar limit is never attained for a nonzero value of $\alpha$ and large values of central Fermi momentum.  One can truly obtain Chandrasekhar's limiting mass by completely neglecting the effect of quantum fluctuations of spacetime by setting $\alpha=0$ also shown in Figure \ref{Figure2}a. Thus the noncommutative situation  is quite unlike the standard theory of white dwarfs where one can reach the Chandrasekhar mass in the limit $\xi_c\rightarrow\infty$. 

Although the deformation parameter $\alpha$ is expected to be small, we presume that the effect of Planck scale physics (quantum fluctuation of space-time) provides an effective large-distance description which would alter the dynamics of large-scale systems existing on such backgrounds. Moreover, it is difficult to tackle numerical values with high precision for very small values of $\alpha$. Consequently, to assess the effect, we take  $\alpha=0.01$, the result of which is shown in Figure \ref{Figure3}. For low values of $\xi_c$, it is observed from the right-hand part of Figure \ref{Figure3} that the mass-radius curve approaches the asymptotic behavior $M/M_\odot = 4.6475 (1+2\alpha)^3 (R_0/R)^3$ as obtained from Eqs.~(\ref{eq:MLxi}) and (\ref{eq:RLxi}), which is also shown on the right hand part of the figure.  As $\xi_c$ is increased, the mass increases slowly and the radius decreases, reaching a minimum value $\approx 326$ km,  as can be seen in the inset of Figure \ref{Figure3}. On further increasing  $\xi_c$, the mass and radius both increase boundlessly and behave similar to the asymptotic expression $M/M_\odot = 0.04628 \ (R/\alpha R_0)^3$ obtained from Eqs.~(\ref{eq:MHxi}) and (\ref{eq:RHxi}), as shown in the inset of Figure \ref{Figure3}. However for large $\xi_c$, the asymptotic expression does not coincide exactly with the exact solution because the exact solution has a core of approximately uniform density and the density falls off outside this core whereas the asymptotic was based on the approximate Lane-Emden equation of order zero implying a constant density throughout the star.  These features are displayed in Table \ref{Table:1} where it is shown that the asymptotic values are in better agreement with the exact ones for low values of $\xi_c$ than for high values of $\xi_c$.

\begin{table*}
\caption{\label{Table:1}Masses and radii of Helium white dwarfs with noncommutative equation of state for $\alpha=0.01$. The columns marked ``Asymptotic" correspond to the approximate Eqs.~(\ref{eq:MLxi}), (\ref{eq:RLxi}), (\ref{eq:MHxi}), and (\ref{eq:RHxi}). The columns marked ``Exact" represent the exact solutions of Eqs.~(\ref{eq:FHS1}) and (\ref{eq:FHS2}).}
\centering
\begin{tabular}{cccccccccc}
\hline
Low&\multicolumn{2}{c}{Asymptotic} &\multicolumn{2}{c}{Exact}  &High&\multicolumn{2}{c}{Asymptotic}&\multicolumn{2}{c}{Exact}\\
\hline
$\xi_c$ &$R$ \ [km]& $M$ \ [M$_{\odot}$]&$R$ \ [km]& $M$ \ [M$_{\odot}$] & $\xi_c$ &$R$ \ [km]& $M$ \ [M$_{\odot}$]&$R$\  [km]& $M$ \ [M$_{\odot}$] \\ 
\hline
$0.09$  &  $33786.90$ & $0.0193$ & $33777.13$ & $0.0192$ & $95.0$ & $414.76$ & $292.7555$ & $473.68$ & $242.6950$\\
$0.10$  & $32053.07$  & $0.0225$ & $32041.25$ & $0.0225$&$96.0$   & $466.15$ & $415.6153$ & $518.51$ & $350.0674$\\  
$0.11$  & $30561.40$  & $0.0260$  & $30547.46$ & $0.0259$ &$97.0$  & $541.06$ & $649.9058$ & $586.17$ & $558.8757$\\  
$0.12$  & $29260.31$  & $0.0296$  & $29244.17$ & $0.0295$&$98.0$   & $666.07$ & $1212.4638$& $702.76$ & $1072.0714$\\ 
$0.13$  & $28112.40$  & $0.0334$   & $28093.96$ & $0.03321$&$99.0$ & $946.79$ & $3481.9895$& $972.68$ & $3205.1357$\\
 \hline
\end{tabular}
\end{table*}

Although these conclusions are based on not very small value of $\alpha$ ($=0.01$), we expect the same qualitative behavior for lower values of $\alpha$. This is in fact clear from the mass-radius curves shown in Figure \ref{Figure2}b where the mass is seen to diverge for very large values of $\xi_c$ even for the case $\alpha=10^{-4}$. To get an approximate idea, we calculate mass and radius values from the asymptotic relations  given by Eqs.~(\ref{eq:MHxi}) and (\ref{eq:RHxi}) with central Fermi momentum close to $\xi_{\rm max}$. If we take $\xi_c=(1-\delta)\xi_{\rm max}=(1-\delta)/\alpha$, then the asymptotic values of mass and radius turn out to be $M=\{18(1-\delta)/\delta\}^{3/2} M_0/9$ and $R=\alpha \{18(1-\delta)/\delta\}^{1/2} R_0$. Table \ref{Table:B} demonstrates that the mass and radius values can be excessively large when the central Fermi momentum $\xi_c$ approaches $\xi_{\rm max}$ sufficiently closely, when $\alpha$ is made very small.

Thus for very low values of $\alpha$ we do not expect the Chandrasekar limit even when the central Fermi momentum is taken to be very large. This behavior is in contrast with  the ideal (commutative) case where the radius decreases to zero and the mass increases and approaches the Chandrasekar limit as $\xi_c\rightarrow\infty$. Thus it suggests that quantum space-time fluctuations play a significant role in determining the mass-radius relation.

\section{Limitation due to Neutronization}\label{N}

The preceding analysis  suggests that the inclusion of quantum space-time fluctuations (via a noncommutative geometry) in the dispersion relation and hence into the equation of state affects the existence of the Chandrasekhar limit significantly. It predicts white dwarfs with mass exceedingly larger than the Chandrasekhar mass with large radii. This obviously is in disagreement with observed non-magnetic white dwarfs that are found only in the mass range $0.17$ M$_\odot$$-$$1.33$ M$_\odot$ \cite{Shipman1972,Shipman1977,Shipman1979,Vennes1997, Marsh1997,Vennes1999, Kilic2007}. In this section, we propose a realistic model of white dwarfs by including neutronization which can actually resolve these difficulties. 

\begin{table*}
\caption{\label{Table:B}Asymptotic values of masses and radii of Helium white dwarfs with noncommutative equation of state following from Eqs.~(\ref{eq:MHxi}) and (\ref{eq:RHxi}) when $\xi_c$ is close to $\xi_{\rm max}$ such that $\xi_c=(1-\delta)\xi_{\rm max}$, with $\delta$ small.}
\centering
\begin{tabular}{ccccccccc}
\hline
\multicolumn{3}{c}{$\alpha=1.0\times10^{-05}$} &\multicolumn{3}{c}{$\alpha=3.0\times10^{-10}$}&\multicolumn{3}{c}{$\alpha=5.0\times10^{-22}$}	\\	  
$\delta$            & $R$ \ [km]            & $M$ \ [M$_\odot$]   & $\delta$            & $R$ \ [km]             & $M$ \ [M$_\odot$]    & $\delta$            & $R$ \ [km]             & $M$ \ [M$_\odot$]    \\
\hline
$10^{-6}$ & $9.4\times10^{01}$ & $3.1\times10^{10}$ & $10^{-12}$  & $2.9\times10^{00}$  & $3.1\times10^{19}$ &$10^{-30}$ & $4.7\times10^{-3}$  & $3.1\times10^{46}$  \\  
$10^{-12}$ &  $9.4\times10^{04}$ & $3.1\times10^{19}$&$10^{-18}$ & $2.9\times10^{04}$ & $3.1\times10^{28}$ &$10^{-40}$  & $4.7\times10^{02}$  & $3.1\times10^{61}$   \\  
$10^{-18}$ & $9.4\times10^{07}$   & $3.1\times10^{28}$&$10^{-26}$ & $2.9\times10^{07}$ & $3.1\times10^{40}$ &$10^{-50}$  & $4.7\times10^{07}$ & $3.1\times10^{76}$         \\ 				           
 \hline
\end{tabular}
\end{table*}

It is well-known that neutronization, or inverse $\beta$-decay ($^A_ZX + e\longrightarrow ^{\ \ A}_{Z-1}\!\!Y +\nu_e$), takes place at a sufficiently high density. Since the density determines the Fermi energy $E_F$, condition of inverse $\beta$-decay is satisfied  when $E_F\geqslant \varepsilon_Z$, where  $\varepsilon_Z$ is the difference in binding energies of the parent and daughter nuclei. Following \cite{Salpeter1961}, we calculate the threshold density $\rho_{\beta}$ by setting $E_F=\varepsilon_Z$ (excluding the electron rest mass) and  obtain $\xi_\beta$ ($=p_\beta/m_e c$) as 
\begin{equation}\label{eq:NT}
\small
\xi_\beta=\frac{\varepsilon_Z}{m_ec^2}\left\{ 1 + 2\frac{m_ec^2}{\varepsilon_Z} \right\}^{1/2} \left\{1+\alpha \left(1+ \frac{\varepsilon_Z}{m_ec^2} \right)\right\}^{-1}
\end{equation}
using the noncommutative dispersion relation given by Eq.~(\ref{eq:dispersion}).
For Helium, $\varepsilon_Z =20.596$ MeV, as obtained from Table \rom{2} of \cite{Rueda2011}.

In the noncommutative framework, the equations for hydrostatic equilibrium are expressed by Eqs.~(\ref{eq:FHS1}) and (\ref{eq:FHS2}). Since Eq.~(\ref{eq:FHS1}) contains the parameter $\alpha$ through its dependence on the noncommutative dispersion function $f(\xi)$, their solution yields different values for different choices of $\alpha$, or equivalently $\lambda$. Consequently, we solve Eqs.~(\ref{eq:FHS1}) and (\ref{eq:FHS2}) numerically for different values of $\alpha$ taking the central value as the neutronization threshold $\xi_\beta$.  It may be noted that $\xi_\beta$ also depends on the choice of the $\alpha$ value through Eq.~(\ref{eq:NT}). The inset of Figure \ref{Figure2}b shows the neutronization points for $\alpha=10^{-2}$, $10^{-3}$, and $10^{-4}$ (triangle, square, and circle, respectively). 

Table \ref{Table:2} gives the values of $\xi_\beta$ corresponding to different deceasing values of $\alpha$, or $\lambda$. It is clear that it is not possible to have an exceedingly large central value $\xi_c$ corresponding to these values of $\alpha$ due to the neutronization threshold.  The mass and radius of white dwarfs obtained via exact solution of Eqs.~(\ref{eq:FHS1}) and (\ref{eq:FHS2}) are shown in the last two columns of Table \ref{Table:2}. We note that both mass and radius take finite values. For large values of $\alpha$, such as $10^{-2}$ (and higher), the mass increase to values higher than the Chandrasekhar mass. However, as the $\alpha$ value is decreased to $10^{-3}$, the mass is $1.5495$ M$_{\odot}$, and the radius $613.2817$ km. On gradually decreasing $\alpha$, we see that the mass appears to approach the limits $1.4518$ M$_{\odot}$ and the radius $601.1821$ km, which are obtained for the case $\alpha=10^{-7}$. This is not a very low value for $\alpha$, because if we speculate that $\lambda\sim1/M_P c^2$, where $M_P=\sqrt{\hbar c/G}$ is the Planck mass, $\alpha=\lambda m_e c^2 \sim m_e/M_P\sim10^{-23}$. We see that the $\alpha=10^{-7}$ values are close to the ideal ($\alpha=0$) values of $1.4518$ M$_{\odot}$ and the radius $601.18$ km for $\xi_\beta=41.2932$. However when we disregard the neutronization threshold for the ideal case and  seek solutions for very large value of $\xi_c$ (beyond $\xi_\beta$) we obtain the Chandrasekhar limiting mass as $1.4562$ M$_{\odot}$ with zero radius. The slight difference from the quoted value of $1.44$ M$_{\odot}$ is because of a slightly different numerical accuracy in our computation.

An estimate of the quantum gravitational parameter $E_{\rm QG}\sim\lambda^{-1}$ (defined in Section \ref{intro}) was obtained from the observed time-dalay in the arrival of TeV-scale photons from $\gamma$-ray flares in a distant galaxy, the AGN Markarian 421. This suggested a lower bound of $E_{\rm QG}\sim10^{16}$ GeV (or $\lambda\sim10^{-20}$ MeV$^{-1}$) \cite{Biller1999}. On the other hand, a value of $E_{\rm QG}\sim10^{18}$ GeV (or $\lambda\sim10^{-21}$ MeV$^{-1}$), was suggested from the compatibility between data obtained from AGN Markarian 501 and PKS $2155$-$304$ \cite{Ellis2009}. This latter value of $\lambda$ gives $\alpha\sim5\times10^{-22}$, suggesting the limit $\alpha\ll10^{-7}$.   

We see from Table \ref{Table:2} that $\alpha\ll  \xi_\beta$ for low values of $\alpha$ and we thus make use of this limit to solve Eqs.~(\ref{eq:FHS1}) and (\ref{eq:FHS2}) approximately. Eq.~(\ref{eq:f'}) gives $f'(\xi)\approx1$ for extremely low values of $\alpha$, so that Eq.~(\ref{eq:mylaneemdeneq}) approximates to 
\begin{equation}\label{eq:Emden3}
\frac{1}{\eta^2} \frac{d}{d\eta}  \left(\eta^2 \frac{d\theta }{d\eta}      \right)+\theta^3 = 0,
\end{equation}
where $\theta=\xi/\xi_c$ and $\eta=\xi_c \ x/\sqrt{3}$ is a dimensionless radius. This equation is solved with boundary conditions $\theta(0)=1$ and $\theta(\eta_R)=0$, where $\eta_R$ corresponds to the radius of the white dwarf. Eq.~(\ref{eq:Emden3}) is the Lane-Emden equation of index 3, whose numerical solution is given in \cite{Weinberg1972}. 

From Eq.~(\ref{mass_equation}) we obtain the dimensionless mass and radius and using Eq.~(\ref{eq:NT}) in the limit $\alpha\ll10^{-7}$, we obtain the dimensionless radius of the white dwarf as 
\begin{equation}\label{eq:neutro}
\small
\tilde{M}_\beta = \sqrt{3} \left(-\eta^2 \frac{d\theta}{d\eta}\right)_{ \eta=\eta_R}, \hspace{0.2cm} \tilde{R}_\beta=\frac{m_ec^2}{\varepsilon_Z}\sqrt{ \frac{3}{1+\frac{m_ec^2}{\varepsilon_Z}}} \eta_R
\end{equation}
giving the mass as $M_\beta=M_0\tilde{M}_\beta=1.4563 \ \rm{M}_{\odot}$ and the radius as $R_\beta=R_0\tilde{R}_\beta=648.809$ km, using $\eta_R=6.89685$ and $-\eta^2 \left(d\theta/d\eta\right)_{ \eta=\eta_R}= 2.01824$ for $n=3$. These mass and radius values, being approximate, do not coincide with the numerical solutions given in the last few rows of \ref{Table:2}. We also note that for values of $\xi_c$ lower than the $\xi_\beta$, lower values of masses with higher values of radii are possible solutions (for any value of $\alpha$) as shown in the right-hand part of Figure \ref{Figure2}b. 

 \begingroup
\squeezetable
\begin{table}
\caption{\label{Table:2}Masses and radii of Helium white dwarfs with the non commutative equation of state for different values of $\alpha$ when the central Fermi momentum is taken to be the neutronization threshold $\xi_\beta$, Eq.~(\ref{eq:NT}), the corresponding neutronization density $\rho_\beta$ is given by Eq.~(\ref{eq:density}). The displayed results represent exact solutions of the equation of hydrostatic equilibrium, Eqs.~(\ref{eq:FHS1}) and (\ref{eq:FHS2}).}
\centering
\begin{tabular}{cccccc}
\hline
$\alpha$   &  $\lambda$\ [MeV$^{-1}]$  & $\xi_\beta$ & $\rho_\beta$ \ [g/cm$^3$]   &$R_\beta$ \ [km] & $M_\beta$  \ [M$_{\odot}$] \\
\hline
$2.0\times10^{-2}$            & $3.91\times10^{-2}$                 & $22.613$        &    $2.252\times10^{10}$      & $948.55$             & $3.9954$        \\  
$1.0\times10^{-2}$            & $1.96\times10^{-2}$                 & $29.223$        &    $4.861\times10^{10}$      & $753.24$             & $2.5736$          \\  
$1.0\times10^{-3}$            & $1.96\times10^{-3}$                 & $39.655$         &    $1.214\times10^{11}$     & $613.28$             & $1.5495$       \\  
$1.0\times10^{-4}$            & $1.96\times10^{-4}$                 & $41.123$         &    $1.355\times10^{11}$       & $602.35$             & $1.4614$      \\
$1.0\times10^{-5}$            & $1.96\times10^{-5}$                 & $41.276$         &    $1.370\times10^{11}$       & $601.30$             & $1.4527$      \\ 
$1.0\times10^{-7}$            & $1.96\times10^{-7}$                 & $41.293$         &    $1.371\times10^{11}$       & $601.18$             & $1.4518$       \\
 \hline
\end{tabular}
\end{table}
 \endgroup
 
 As noted earlier in Tables \ref{Table:B}, the noncommutative equation of state allows for extremely high values of mass and radius if the effect of neutronization is neglected so that the central Fermi momentum could approach $\xi_{\rm max}=1/\alpha$. However, due to the constraint of neutronization,  the $\xi_c$ value cannot  approach a value higher than $\xi_\beta$. Together with the neutronization constraint on $\xi_c$, when we take $\alpha\ll10^{-7}$ as suggested by observations from $\gamma$-ray burst, the mass and radius values approach finite values as we have seen in Table \ref{Table:2}.

 \begin{table*}
\caption{\label{Table:3}Masses and radii of Carbon and Oxygen white dwarfs with the noncommutative equation of state for different values of $\alpha$ when the central Fermi momentum is taken to be the neutronization threshold $\xi_\beta$, Eq.~(\ref{eq:NT}), the corresponding neutronization density $\rho_\beta$ is given by Eq.~(\ref{eq:density}). The displayed values of masses and radii represent exact solutions of the equation of hydrostatic equilibrium, Eqs.~(\ref{eq:FHS1}) and (\ref{eq:FHS2}).}
\centering
\begin{tabular}{ccccccccc}
\hline
&\multicolumn{4}{c}{$\prescript{12}{6}{\mathbf{C}}$, \ $\varepsilon_Z=13.370$  \ MeV} &\multicolumn{4}{c}{$\prescript{16}{8}{\mathbf{O}}$, \   $\varepsilon_Z=10.419$ \ MeV } \\
$\alpha$    & $\xi_\beta$   &$\rho_\beta$ \ [g/cm$^3$] &$R_\beta$ \ [km] & $M_\beta$ \ [M$_{\odot}$] & $\xi_\beta$   & $\rho_\beta$ \ [g/cm$^3$] &$R_\beta$ \ [km] & $M_\beta$ \ [M$_{\odot}$] \\
\hline
$2.0\times10^{-2}$            &  $17.590$   &$1.060\times10^{10}$&    $1206.20$  &$2.9795$   &  $14.964$   &$6.527\times10^{09}$&    $1406.70$  &$2.5994$        \\  
$1.0\times10^{-2}$            & $21.347$     &$1.895\times10^{10}$  & $1028.13$  &   $2.1463$    & $17.601$  &$1.062\times10^{10}$& $1236.10$   &  $1.9783$      \\  
$1.0\times10^{-3}$            & $26.428$  & $3.595\times10^{10}$  &$895.22$    &   $1.5100$    & $20.919$  &$1.783\times10^{10}$& $1105.41$   & $1.4903$       \\  
$1.0\times10^{-4}$            & $27.072$   & $3.865\times10^{10}$  &$883.98$   &   $1.4525$    & $21.320$  & $1.888\times10^{10}$&$1093.92$  & $1.4453$         \\
$1.0\times10^{-5}$             & $27.139$   & $3.893\times10^{10}$  &$882.89$    & $1.4468$      & $21.361$  & $1.898\times10^{10}$&$1092.79$   & $1.4409$\\ 
$1.0\times10^{-7}$            & $27.146$   &  $3.896\times10^{10}$  &$882.77$    & $1.4462$      &$21.366$   &$1.899\times10^{10}$&$1092.67$   &  $1.4404$\\
 \hline
\end{tabular}
\end{table*}

\section{Discussion and Conclusion}\label{conclusion} 

The effect of quantum gravity, although very small, is inevitably  present everywhere. We thus expect that it will modify the stability of astrophysical objects such as white dwarfs. In particular, it is already well-known that the noncommutative formulation of quantum gravity modifies the dispersion relation of any particle. To study the effect of such modification on the stability of white dwarfs, we employed a modified dispersion relation of the form $E_{\bf p}^2=\mathbf{p}^2c^2(1+\lambda E_{\bf p})^2+m^2c^4$ and observed that the equation of state of a degenerate electron gas undergoes a substantial modification as a result of the emergence of a cutoff momentum $p_{\rm max}= 1/\lambda c$ inherent in the dispersion relation. As a consequence, the possible values of masses and radii of white dwarfs change from the ideal case.

We have analyzed the situation in two different ways in the framework of Newtonian gravity. First, we employed the equation of hydrostatic equilibrium to obtain an approximation in the limit of low central Fermi momentum resulting in the Lane-Emden equation of index $3/2$. On analyzing the solutions we found that both the mass and radius are affected by the parameter $\lambda$ indicating the persistence of the effect of noncommutative equation of state even for low mass white dwarfs. Next, we analyzed the problem when the central Fermi momentum $p_{\rm Fc}$ approaches $p_{\rm max}=1/\lambda c$. On working out the asymptotics, the Lane-Emden equation of index zero is obtained that clearly indicated that both mass and radius would approach infinity when the central Fermi momentum approaches $p_{\rm max}=1/\lambda c$. 

Finally, we solved the equations of hydrostatic equilibrium exactly by numerical means with the noncommutative equation of state without making any approximations to the modified dispersion relation. We found that  the modified mass-radius curve did not coincide with the ideal degenerate curve even in the low central Fermi momentum region. This can be associated with the previous asymptotic solution in the low momentum limit where the masses and radii were found to have small departures in terms of the noncommutative parameter $\lambda$. On the other hand, we observed a strong departure from the ideal mass-radius curve for high values of the central Fermi momentum even for a low value such as $\alpha = \lambda m_ec^2=10^{-4}$. This trend is expected to be qualitatively the same for values of $\alpha$ lower that $10^{-4}$. Since it is difficult to handle very low values of $\alpha$ numerically, we assessed the situation for a few higher values of $\alpha$ such as $10^{-1}$, $10^{-2}$, $10^{-3}$, and $10^{-4}$. For all these cases, we found masses excessively larger than the Chandrasekhar bound. The approach to high mass values is delayed (with respect to increase in central Fermi momentum) when the $\alpha$ value is decreased. It was clear that even for lower values of $\alpha$, the mass would approach very large values higher than the Chandrasekhar limit with large values of radii. This was confirmed from our asymptotic analysis for any low magnitude of $\alpha$ when $\xi_c$ approaches $\xi_{\rm max}$.

The above situation is remarkably different from observations on non-magnetic white dwarfs because they are found in the mass range from $0.17$ M$_{\odot}$ \cite{Kilic2007} to $1.33$ M$_{\odot}$ \cite{Vennes1997, Vennes1999, Marsh1997, Kepler2007} with radii ranging from $0.0153$ $R_{\odot}$ ($10644$ km) to $0.0071$ $R_{\odot}$ ($4939$ km) \cite{Shipman1972, Shipman1977, Shipman1979}. This disagreement can be reconciliated by noting that the central Fermi momentum of white dwarfs cannot take arbitrarily high values as it is limited by the neutronization threshold. In addition, the quantum gravity parameter $\lambda$ is also not large. Consequently we solved the equations of hydrostatic equilibrium with a few values of $\alpha$ ranging from $10^{-2}$ to $10^{-7}$ with the central Fermi momentum taken as the neutronization threshold. Although the case $\alpha=10^{-2}$ yielded a mass as large as $2.5736$ M$_\odot$ for Helium, as the $\alpha$ value is decreased to $10^{-5}$, we found the mass as $1.4527$ M$_\odot$ with a radius $601.29$ km. On further deceasing the $\alpha$ value these values did not change appreciably.
 
 The parameter $\lambda$ may be inversely proportional to the the quantum gravity parameter $E_{QG}$ that occurs in the dispersion relation  $c^2p^2=E^2[1+\sigma E/E_{QG} + \ldots]$ for propagation of photons through vacuum. In particular, \cite{Ellis2009} predicted the time delay in receiving $\gamma$-ray photons from distant active galaxies that suggested the value $E_{QG}\sim10^{18}$ GeV. This value of $E_{QG}$ translates to $\alpha\sim5\times10^{-22}$ if we assume $\lambda\sim E_{QG}^{-1}$. We thus expect that the $\alpha$ value to be lower than $10^{-5}$. In our numerical calculation, when we decreased the $\alpha$ value from $10^{-4}$ to $10^{-7}$, we saw that the mass and radius approach the limiting values of $1.45$ M$_\odot$ and $601$ km (for Helium) at the neutronization threshold. For values of the central Fermi momentum lower than the neutronization threshold, we obtained lower values of masses with larger values of radii. 
 
 It may however be noted that the above observations are for photons from $\gamma$-ray bursts propagating through vacuum. Equivalent data for massive particles, such as electrons, do not exist in the literature and it is difficult to guess the value of $\alpha$ for electrons. Consequently, we have shown the neutronization threshold values for the masses and radii of  some white dwarfs ($\prescript{4}{2}{\mathbf{He}}$, $\prescript{12}{6}{\mathbf{C}}$, and $\prescript{16}{8}{\mathbf{O}}$) for values of $\alpha$ ranging from $2.0\times10^{-2}$ to $1.0\times10^{-7}$ in Tables \ref{Table:2} and \ref{Table:3}.  We note that the neutronization threshold value for a pure $\prescript{16}{8}{\mathbf{O}}$ white dwarf should be the same as that of a carbon-oxygen white dwarf because the core of the latter is expected to be purely $\prescript{16}{8}{\mathbf{O}}$ and neutronization is expected to begin at the center \cite{Canal1980}. The top row of Table  \ref{Table:3} for $\alpha=2.0\times10^{-2}$ indicates that a carbon-oxygen white dwarf would have a critical mass of $2.5994$ M$_\odot$. There have been a few observations of overluminous type Ia SNe (SN 2003fg, SN 2006gz, SN 2007if, SN 2009dc) \cite{Howell2006, Hicken2007, Yamanaka2009, Scalzo2010, Silverman2011} that produced a high amount of  $\prescript{56}{}{\mathbf{Ni}}$ ranging from $1.2$ M$_\odot$ to $1.7$ M$_\odot$ suggesting their progenitors to be super-Chandrasekhar white dwarfs ranging from $2.2$ M$_\odot$ to $2.8$ M$_\odot$.
 
However, \cite{Hicken2007} argued that the type Ia SN 2006gz was a double degenerate (DD) merger of two sub-Chandrasekar white dwarfs as supported by the unusually low and slowly declining Silicon velocity which is also predicted by DD models. \cite{Silverman2011} speculated that SN 2009dc was very likely due to the merger of two white dwarfs as supported by simulations. \cite{Chen2009} considered a single-degenerate white dwarf supported by differential rotation accreting at a low rate from a normal companion.  With an initial $1.2$ M$_\odot$, they found the possibility of having a super-Chandrasekhar SNe Ia event. However, white dwarfs with mass exceeding $1.7$ M$_\odot$ was predicted to be not likely. \cite{Das2012} indicated that the presence of a strong magnetic field ($\sim10^{15}$ Gauss) in a white dwarf can support a mass of 2.3--2.6 M$_\odot$ due to the role of Landau levels. On the other hand, pointing to various disagreements among the existing SNe Ia models,  \cite{Kerkwijk2013} argued that SNe Ia events generally happen due to the merger of two carbon-oxygen white dwarfs.
 
 Thus it appears that the super-Chandrasekhar scenario is not possible in the case of a normal white dwarf (without rotation or magnetic field). We are thus led to infer that it is neutronization that would constrain white dwarfs within the Chandrasekhar limit. This implies that $\alpha$ should be very small ($\sim10^{-7}$ or lower) so that a mass close to the Chandrasekhar limit is obtained as a consequence of the neutronization threshold. If this was not the case, that is, in the absence of neutronization, the modified dispersion relation would support excessively high mass values (beyond the Chandrasekar mass) even for very low values of $\alpha$ such as $10^{-7}$ or lower. It is only when we impose the condition of neutronization (on the top of the effect of quantum gravity) that we get mass limits close to the Chandrasekhar mass.

We further note that since the effect of quantum gravity must be inevitably present, we should consider it in the analysis. There are two simple ways to take the effect of quantum gravity into account. One way is to take this effect through noncommutativity that modifies the dispersion relation as presented in this paper. Another way is to take this effect through a generalized uncertainty relation as discussed earlier in \cite{Mathew2018}. Based on these two differing approaches, we may state that whichever way we attempt to include the effect of quantum gravity in the description, we find that white dwarfs with excessively high masses  (beyond the Chandrasekar mass) would be supported although the quantum gravity parameter is taken to be extremely small. In both approaches, we find that it is only when we impose the condition of neutronization that we obtain mass limits close to the Chandrasekar mass. Thus, in realistic situations, such mass limits exist because of the neutronization threshold that destabilizes the white dwarf due to the onset of inverse $\beta$-decay.

Since the above discussion applies to white dwarfs when the equilibrium is governed by Newtonian gravity, the situation is expected to alter when general relativity is employed for the hydrostatic equilibrium. It is already known for $\prescript{4}{2}{\mathbf{He}}$ and $\prescript{12}{6}{\mathbf{C}}$ white dwarfs that the gravitational instability sets in before the neutronization instability can set in, whereas, for $\prescript{16}{8}{\mathbf{O}}$ white dwarfs, it is the instability due to neutronization that sets in before the gravitational instability. However, in the present case of noncommutative equation of state, the neutronization threshold depends on the noncommutative parameter $\alpha$ according to Eq.~(\ref{eq:NT}). Consequently, it would be interesting to analyze the problem in the general relativistic framework to see to what extent the above situation changes in determining the competition between the two kinds of instabilities. We leave this motivation as an interesting research problem for the future. 

\begin{acknowledgements}
Arun Mathew is indebted to the Ministry of Human Resource Development, Government of India, for financial support through a doctoral fellowship. 
\end{acknowledgements}


%

\end{document}